\documentclass[12pt,preprint]{aastex}

\newcommand{\etal}{et al.~}
\newcommand{\kms}{km s$^{-1}$~}
\newcommand{\vp}{$\upsilon_{p}$}

\shorttitle{The Globular Cluster System of NGC 4636}
\shortauthors{Park et al.}

\begin{document}

\title{The Globular Cluster System of the Virgo Giant Elliptical Galaxy NGC 4636:
    I. Subaru/FOCAS Spectroscopy and Database$^*$}

\author{ Hong Soo Park\altaffilmark{1}, Myung Gyoon Lee\altaffilmark{1}, 
Ho Seong Hwang\altaffilmark{1,2}, Nobuo Arimoto\altaffilmark{3}, Naoyuki Tamura\altaffilmark{4},
and Masato Onodera\altaffilmark{2} }

\email{hspark@astro.snu.ac.kr, mglee@astro.snu.ac.kr, hoseong.hwang@cea.fr,
arimoto.n@nao.ac.jp, naoyuki@subaru.naoj.org, masato.onodera@cea.fr}

\altaffiltext{1}{Astronomy Program, Department of Physics and Astronomy, Seoul National University, Korea}
\altaffiltext{2}{CEA, Laboratoire AIM, Irfu/SAp, F-91191 Gif-sur-Yvette, France}
\altaffiltext{3}{National Astronomical Observatory of Japan, Tokyo, Japan}
\altaffiltext{4}{Subaru Telescope, National Astronomical Observatory of Japan, Hilo, USA}
\altaffiltext{*}{Based on data collected at Subaru Telescope, which is operated by the National Astronomical Observatory of Japan.} 

  \begin{abstract}
We present a spectroscopic study of the globular clusters (GCs) in the giant elliptical galaxy  NGC 4636 
in the Virgo cluster.
We selected target GC candidates using the Washington photometry
derived from the deep CCD images taken at the KPNO 4m.
Then we obtained the spectra of 164 target objects in the field of NGC 4636 using 
the Multi-Object Spectroscopy (MOS) mode of Faint Object Camera and
Spectrograph (FOCAS) on the SUBARU 8.2m Telescope.
We have measured the velocities for 122 objects:
105 GCs  in NGC 4636, the nucleus of NGC 4636, 
11 foreground stars, 2 background galaxies, and 3 probable intracluster GCs  in 
the Virgo cluster.
The GCs in NGC 4636 are located in the projected galactocentric radius within $10\arcmin$
(corresponding to 43 kpc). 
The measured velocities for the GCs range from $\sim 300$ \kms
to $\sim 1600$ \kms, with a mean value of $932_{-22}^{+25}$ \kms, which is in good agreement
with the velocity for the nucleus of NGC 4636, $928\pm 45$ \kms.
The velocity dispersion of the GCs in NGC 4636 is derived to be $231_{-17}^{+15}$ \kms and
the velocity dispersion of the blue GCs is slightly larger than that of the red GCs. 
Combining our results with data in the literature, 
we produce a master catalog of radial velocities for 238 GCs in NGC 4636.
The velocity dispersion of the GCs in the master catalog
is found to be
$225_{-9}^{+12}$ \kms for the entire sample,
$251_{-12}^{+18}$ \kms for 108 blue GCs, and
$205_{-13}^{+11}$ \kms for 130 red GCs.
\end{abstract}

%% Keywords should appear after the \end{abstract} command. The uncommented
 % --- galaxies:spectroscopy}

\section{Introduction}

Globular clusters  (GCs) are invaluable fossils  
that provide critical clues for understanding the formation and
early evolution of their host galaxies.
GCs are found from the central region to the outer halo in their host galaxies, 
and thousands of them are often found in nearby giant elliptical galaxies (gEs).
Thus GCs are an excellent tool to study the structure and kinematics
of the halo of nearby gEs.

While there are numerous studies based on the photometry of the GCs in gEs
(e.g., see the reviews by \citealt{lee03,bro06}),
  there are only a small number of studies based on the spectroscopy of the GCs in gEs :
M49 \citep{zep00,cot03}, 
M60 \citep{pie06, bri06,lee08a,hwa08}, 
M87 \citep{coh97,kis98b,cot01}, 
NGC 4636 \citep{sch06}, %, cha08}, 
NGC 1399 \citep{kis98, min98, kis99, ric04, ric08}, 
NGC 5128 \citep{pen04a, pen04b, woo07}, and
NGC 1407 \citep{rom09}.
Recent results on the kinematics of the GCs in gEs are summarized as follows.

\citet{cot01} presented a kinematic study of 278 GCs in M87 located close to the dynamical center of the Virgo Cluster.
All the GCs, the blue GCs, and the red GCs appear to rotate with similar rotation amplitudes.
However the blue GCs rotate around the photometric minor axis in the outer region ( $R>2 R_{\rm eff}$),
while they appear to rotate around the photometric major axis in the inner region ( $R<R_{\rm eff}$).
The entire GC system has an almost perfectly isotropic velocity ellipsoid,
The blue GC orbits are modestly tangentially biased,
while the red GG orbits are radially biased.

\citet{cot03} presented a kinematic study of 263 GCs in M49, the brightest galaxy in Virgo,
finding that the GC system of M49 exhibits a slow rotation
that is mainly due to a net rotation of the blue GCs alone.
The red GCs show essentially no rotation, but
they show a weak rotation beyond 0.5$R_{\rm eff}$ in the opposite
direction to that of the blue GCs.
In the outer region ( $R \gtrsim R_{\rm eff}$) of M49, the velocity dispersion
for the blue GCs exceeds that for the red GCs by $\sim 50\%$.
The entire GC system in M49 is  consistent with an almost perfectly isotropic 
velocity ellipsoid.

\citet{ric04} presented a kinematic study of 468 GCs in NGC 1399 located at the center of the Fornax cluster.
The velocity dispersion for the red GCs is smaller than those for the blue GCs.
Little rotation is found for either of the blue or red GCs.
The red GC orbits are isotropic, and the blue GCs are slightly tangentially biased.

\citet{woo07} presented a kinematic study of 336 GCs in NGC 5128.
The kinematics of %158 
the red GCs and %178 
blue GCs are quite similar.
The GCs in the inner region (within 5 kpc) have a lower rotation signal 
than those in the outer regions.
The velocity dispersions for both the blue and red GCs display a steady increase with galactocentric radius.

\citet{hwa08} presented a kinematic study of 121 GCs in M60, a gE in Virgo.
The velocity dispersion of the red GCs is marginally larger than that of the blue GCs.
The GC system shows a significant overall rotation,
and the rotation of the blue GCs is slightly weaker than or similar to that of the red GCs.
The GC system is found to have a tangentially biased velocity ellipsoid.
The blue GC orbits are modestly tangentially biased, 
while the red GC orbits are modestly radially biased.

\citet{rom09} presented a kinematic study of 172 GCs in NGC 1407.
They found a weak rotation in the outer region of NGC 1407, 
and a rotational misalignment between the blue and red GCs.
The GC system has somewhat tangential anisotropy, which is mainly due to the blue GCs.
The red GC orbits are almost isotropic.

%%%
These previous studies showed that
the kinematics of the GC systems in these gEs is diverse, showing a large difference in velocity dispersion, rotation, and their radial variation. 
It is not yet known whether this kinematic diversity is an intrinsic property, or
is due to a small number of GCs in each gE or a small number of  gEs in previous studies.
It is needed to increase the sample size of the GCs in these gEs and
to extend the kinematic study of the GC systems to more galaxies.

We have been carrying a project to investigate the spectroscopic properties of the GCs in gEs. %nearby galaxies. 
We presented our study on the kinematics of the GC system in M60 in Virgo in \citet{lee08a} and \citet{hwa08}.
%, that on the M31 GC system in \citet{lee08c}. 
In this paper and its companion paper \citep{lee09} we present  a study of the kinematics of the GC system in  another gE, NGC 4636, in Virgo.

NGC 4636 
is located $10^{\circ}.8$ (about 2.8 Mpc) south east from the Virgo center, and is considered to be a
major member of a small group falling into the Virgo center \citep{nol93}.
NGC 4636 is  relatively less luminous ($M_V=-21.7$ mag) among the gEs in Virgo, but
it shows several interesting features.
First, it is one of the most X-ray luminous elliptical galaxies in the nearby universe, with $L_X \sim 2 \times 10^{41}$ ergs s$^{-1}$\citep{for85}. It shows a large diffuse X-ray emission extending out to about $10\arcmin$, arm-like asymmetric features involved with shocks in the central region,  and a few hundreds of point sources in the X-ray images\citep{for85,mat98,jon02,loe03,osu05,kim06,pos09}.
%Second, it shows dust lanes emitting mid-infrared(MIR) and far-infrared (FIR) emission in the central region\citep{tem03,tem07}. 
Second, the color distribution of most GCs is bimodal, 
  while the brightest GCs have an intermediate color \citep{dir05}.
  These results were found from Washington photometry of a $34'.7 \times 34'.7$ field obtained 
  using the MOSAIC wide-field CCD camera at the CTIO 4m telescope.
Third, its specific frequency of GCs,
$S_N$=$6\sim 9$, is known to be much higher than that of normal elliptical galaxies \citep{kis94,dir05}.
Fourth, \citet{loe03} and \citet{cha08} suggested that the fraction of dark matter in this galaxy is much larger compared with other galaxies, while \citet{sch06} claimed from the study of kinematics of the GC system in this galaxy that it is only 20 to 30 \%. 
Finally, it has a kinematically decoupled core showing an irregular velocity curve in the central region \citep{cao00}.

 We have been doing a spectroscopic study of the GCs in NGC 4636
using the spectra obtained at the 8.2m Subaru Telescope.
We present the measurement of radial velocities for the GCs in NGC 4636 in this paper,
and will present the kinematic analysis of the data in the companion paper \citep{lee09}.  
%(Lee et al. 2009, in preparation). %\citep{lee09}. %par09a}.
During our study, 
Schuberth et al. (2006, S06) %\citet{sch06} 
presented a kinematic study of 174 GCs in NGC 4636, using the
radial velocity data obtained at the VLT.
Our study is complementary to this study.

We adopted a distance to NGC 4636, % 16.83 Mpc ($(m-M)_0=31.13\pm 0.15$)  ton01},
14.7 Mpc [$(m-M)_0=30.83\pm 0.13$], as given by \citet{ton01} %Tonry et al. (2001)  
based on the surface brightness fluctuation method, % in \citet{ton01},  %mei07}
for which one arcsec in the sky corresponds to 71 pc. %71.05      84 pc. %81 pc.
Foreground reddening toward NGC 4636 is very small, $E(B-V)=0.028$ \citep{sch98},
corresponding to $E(C-T_1)=1.966 E(B-V)=0.055$ and $A(T_1)=0.075$ and $A(V)=0.092$.
Effective radius, ellipticity, and position angle of NGC 4636 are 
$R_{\rm eff}=8.15$ kpc, $\epsilon = 0.22$, and PA$=110$ deg, 
$R_{\rm eff}=1\arcmin.49=6.35$ kpc, $\epsilon_{\rm eff} = 0.256$, and PA$_{\rm eff}=148$ deg, 
$R_{25}=3.60\arcmin=15.34$ kpc, $\epsilon_{25} = 0.326$, and PA$_{25}=148$ deg, 
respectively \citep{kim06,par09}.
%Table 1 lists a list of basic parameter of NGC 4636.

This paper is composed as follows.
Section 2 describes the selection of the targets for spectroscopy,
spectroscopic observation, data reduction, and radial velocity measurement.
In \S3 we present the data for the radial velocities for the objects measured in this study,
and compare our measurements with previous studies. 
We produce a master catalog for the radial velocities for GCs and other objects
in the field of NGC 4636, combining our measurements with the data in the literature.
Using this catalog we select the genuine GCs of NGC 4636. 
Then we show the distribution of the radial velocity of the GCs
with respect to color, magnitude, position angle and galactocentric distance.
Primary results are summarized in the final section.

\section{Observation and Data Reduction}

\subsection{Spectroscopic Target Selection}

We selected the spectroscopic targets using the catalog of GC candidates
obtained from 
Washington $CT_1$ images of a $15.8\arcmin\times15.8\arcmin$ field including the center of NGC 4636 %obtained 
taken at the KPNO 4m telescope.
A photometric study of the GCs in NGC 4636 based on these data will be given in
\citet{par09}. %Park et al. 2009 (in preparation). %  %b}. 
GCs in NGC 4636 appear as point sources in the KPNO images,
and there are some foreground stars and distant background galaxies as well as GCs
among the point sources found in the KPNO images. 
Most of the bright GCs in Virgo gEs have colors $0.9\le(C-T_1)<2.1$
\citep{lee03,dir05,par09}. %, Park et al. 2009, in preparation).
% (\citealt{dir05,par09b}).
We selected, as the spectroscopic targets, the bright GC candidates 
with $0.9\le(C-T_1)<2.1$ and $19<T_1<22.5$ mag.
We also included two small galaxies and the nucleus of NGC 4636 in the list of
targets for comparison.
The number of the targets in the final sample %of globular cluster candidates
is 168.

\subsection{Observation}

Spectroscopic data of 168 targets were obtained using the Multi-Object Spectrograpy (MOS) mode of Faint Object Camera and Spectrograph (FOCAS; \citealt{kas02}) at the Subaru 8.2m Telescope on 2002 April 20 and 21.
A mask covers a circular field of view with diameter of $6\arcmin$. 
We observed five masks, and the positions of the observed masks are shown in Figure 1. 

Figure 1 displays the grayscale map of the $T_1$ image of NGC 4636 taken at the KPNO 4m, showing the positions of the masks.
We created first the model image of the galaxy halo light using the ellipse fitting task IRAF/ELLIPSE\footnotemark\footnotetext
{IRAF is distributed by the
National Optical Astronomy Observatories, which are operated by the
Association of Universities for Research in Astronomy, Inc., under
contract to the National Science Foundation.}.
Then the model image of the halo light of NGC 4636 was subtracted
from the original image to show better the point sources including GCs in NGC 4636, as seen in Figure 1.

Table 1 lists the log of our observation.
We obtained our spectra in the MOS mode with the medium-dispersion blue grism (300B) resulting in a dispersion of 1.34 \AA~ pixel $^{-1}$ 
and the order-cut filter L600 covering $3700-6000$ \AA.
Seeing during the observation was $0.6\arcsec-0.8\arcsec$. 
We obtained mask images of the fields with $R$ filter with exposure times of 180 sec and 300 sec  in the camera mode of FOCAS under the seeing of $\sim1.0\arcsec$ on 2002 March 9. 
Using these images we made masks with  Mask Design Pipeline (MDP), which is a software utility for making masks of MOS \citep{sai03}. 
The slit width along the dispersion axis was 0.8\arcsec~ in designing the MOS mask, giving a spectral resolution of $R \sim 500$.

We obtained three or four 1200 sec exposures for targets in all but one mask field,
with comparison spectra taken with Th-Ar lamps before and/or after each exposure.
For the calibration of the flux, radial velocity and metallicity, 
we used the long-slit spectroscopy mode of FOCAS. 
We observed BD+33d2642, standard star for flux calibration, with 2.0\arcsec~slit width along the dispersion axis and 10 sec exposure time.
Also we observed five Galactic GCs (NGC 5904 (M5), NGC 6205 (M13), NGC 6341 (M92), NGC 6171 (M107), and NGC 6624) for the velocity and metallicity templates, 
with stepping scan mode to sample an area larger than that covered by the slit
by moving the slit along the dispersion direction.
We covered a total area of $6' \times 8'' $ for each cluster.
Basic information of the five Galactic GCs is listed in Table 2 (\citealt{har96} and references therein). The metallicity of these Galactic GCs %globular clusters
ranges from [Fe/H] = $-$2.3 to $-$0.4, and their radial velocity ranges from $v_r= -245.6$ \kms to $+53.9$ \kms.

\subsection{Data Reduction}

We applied basic processing (overscan correction, bias subtraction, and removing cosmic ray) to the CCD images before combining a pair of CCD images using IRAF tasks.
$FORCASRED/bigimage$ task in IDL\citep{sai03} was used for connecting two CCD images and correcting the optics distortion.
Then we flattened multi-slit images with IRAF tasks.
The spectra of the targets from the combined images
were traced, extracted, and sky-subtracted using the IRAF/APALL task. %and  IDENTIFY task.
We could not extract spectra of four %some
faint targets because of low signal-to-noise (S/N) ratio.
We used the Th-Ar lamp spectra for wavelength calibration. 
There are about 40 useful emission lines  in the range of $3800-6000$ \AA. The resulting $rms$ error for wavelength calibration is $\sim 0.8$ \AA.  
Figure 2 represents spectra of two sample GCs in NGC 4636, the nucleus of NGC 4636, and two Galactic GCs. 
Several typical absorption lines for old stellar systems such as G band, H$\beta$ and Mgb are clearly seen in the spectra of Galactic GCs, 
but they are not clearly seen in the spectra of the GCs in NGC 4636 because of low S/N ratio.
The absorption lines such as Mgb and FeI for the red GC in NGC 4636 %and NGC 6624 in the Galaxy
are much stronger than those for the blue GC in NGC 4636, like the Galactic globular clusters
\citep{bro86, coh98, bea00, bro06}.
Absorption features in the spectrum of the NGC 4636 nucleus are much broader than those in the GCs
due to its large velocity dispersion.

\subsection{Velocity Determination}

We determined the radial velocities for the targets
using the Fourier cross-correlation task, IRAF/FXCOR \citep{ton79}.
We fitted the continuum of the spectra using the spline-fit
with $2\sigma$ clipping for low level and $4\sigma$ clipping for
high level, and subtracted the resulting fit from the original spectra.
We used the resulting spectra to measure the radial velocity. % with FXCOR.

%After several tests, 
Since the S/N at $\sim 4000$ \AA~  is  low for most spectra, 
and the strong night sky emission line appears near 5600 \AA,
we decided to use the wavelength range $4200-5400$ \AA~  for cross-correlation.
We measured the radial velocities for the targets,
using five templates of Galactic GCs.
The values derived using four templates (M5, M13, M92 and NGC 6624) are consistent with each other within one $\sigma$, while those derived using M107 show two $\sigma$ difference.
Therefore we took the error-weighted average of these four measurements to get the final value for each target.
The error of the measured radial velocity is then estimated using  
  $<\epsilon_v>=(\Sigma\epsilon_i^{-2})^{-1/2}$.

\section{Results}

\subsection{Velocity Data}

%Among the spectra of 164 targets 
The final number of targets %globular cluster candidates
for which we determined the radial velocities is 122.
We could not determine the radial velocities for 42  objects among the  164 targets because
of poor qualities of the spectra. % with low Signal-to-Noise ratio.
%:105 GCs, 11 foreground stars, 5 small galaxies, and the nucleus of NGC 4636.
The mean S/N of spectra measured the radial velocity is about 13 at $\sim 5000$ \AA.

Figure 3 displays the errors of the measured radial velocities
  versus $T_1$ magnitudes and galactocentric distances of the 119 targets,
  excluding the nucleus of NGC 4636 and two small galaxies having no photometric data.
The errors for our radial velocity measurements %measurements of radial velocities of the targets 
range mostly from 20 to 80 km s$^{-1}$ with a mean of $43\pm18$ km s$^{-1}$.
The mean errors increase with increasing $T_1$ magnitude.
The error spread for the red objects ($(C-T_1 ) \ge1.55$) is similar
to that for the blue objects ($(C-T_1 ) <1.55$), unlike the case seen
in the \citet{sch06} %Schuberth et al. (2006) 
data (their Figure 3) which show a much larger error
spread for the blue objects than for the red objects.
The mean errors of the objects show little dependence on the galactocentric distance.
%and the mean errors for the red objects with $(C-T_1 ) \ge1.55$ are slightly smaller 
%than those for the blue objects with $(C-T_1 ) <1.55$.
%The mean errors  decrease  weakly with the galactocentric distance,
%The mean errors of the bright objects with $T_1<21$ mag are much smaller 
%than those for the faint objects with $T_1\ge21$ mag.
%and those of the bright objects decrease  weakly with the galactocentric distance, 
%while those for the faint objects are indistinct.
%and the mean errors of the bright objects with $T_1<21$ mag are much smaller 
%than those for the faint objects with $T_1\ge21$ mag.

\subsection{Comparison with Previous Studies}

%There is only one previous spectroscopic study of the globular clusters in NGC 4636,  given by \citet{sch06}.
%Schuberth et al. (2006, S06) published a catalog of spectroscopic data including radial velocities for 174 GCs in NGC 4636
\citet{sch06} published a catalog of spectroscopic data including radial velocities for 174 GCs in NGC 4636
based on the spectra obtained using FORS2/XMU at the VLT. %2.
We compared our list with theirs, finding that there are 45 objects in common between the two.
%this study and \citet{sch06}.
We compared our velocity measurements with their measurements
for these common objects, as displayed in Figure 4.
It shows a good agreement between the two studies for most objects, but shows
significant differences for two objects (ID 224 and ID 1049 in this study).
Our values, $880\pm47$ and $1090\pm37$ \kms (for ID 224 and 1049, respectively), are significantly larger than their values, $565\pm16$ and $649\pm40$ \kms.
The reason for these differences for the two objects is not known.
Further spectroscopic observation is needed to check this difference.

From the weighted linear fit to the data discarding these two objects, 
we derive a transformation relation between the two measurements,  with $rms \sim 82$ \kms~,

\begin{equation}
v_{\rm This~ study} = 1.06(\pm0.04)~v_{\rm S06} +1.56(\pm37.68) {\rm~km~s}^{-1}. \label{eq-trans}
\end{equation}
\\
The $rms$ of the relation between the two studies is rather high. % (82 km s$^{-1}$).
The cause for this rather high $rms$ is the existence of some outliers in the common sample between the two studies.
If we use the data without the outliers with more than two times $rms$, 
the $rms$ will be reduced to 61 km/s, 
which is marginally similar to the mean error ($43\pm18$ km s$^{-1}$) of this study.
 
We have determined the radial velocity for the nucleus of NGC 4636 
as $v_p = 928\pm45 $ km~s$^{-1}$.
 This value is consistent with the value, $938\pm4$ km~s$^{-1}$ given by \citet{smi00} %the NED,
and $917\pm31$ \kms in \citet{sch06}. % systemic vel = 906\pm 7km/s by S09
The larger error in our estimate of the nucleus velocity %the NGC 4636 systemic velocity
is primarily due to the fact that we 
used, as templates, the spectra
of GCs with much smaller velocity dispersion to estimate the velocity for
the nucleus of NGC 4636 that has much larger velocity dispersion.

\subsection{A Master Catalog of Velocity Data}

We have produced a master catalog of radial velocities for the objects in NGC 4636,
combining the catalog in this study with the catalog of 174 GCs in NGC 4636 given by \citet{sch06}.
The total number of the objects in the master catalog is 255,
including the nucleus of NGC 4636.
We transformed the radial velocities given by \citet{sch06} to our system,
using the transformation equation (1) %derived in the previous section
for the following analysis.

In Table 3, we list the photometric and kinematic
data set for all 255 spectroscopic targets with measured radial velocities.
The first column represents identification numbers. 
The second and third columns give, respectively, the right ascension and the declination (J2000). 
The galactocentric radius and position angle are given in columns 4 and 5, respectively. 
The magnitude and color information in columns 6 and 7 are from \citet{par09}. 
The eighth column gives the radial velocity and its error measured in this study.
The ninth and tenth columns give the radial velocity (transformed into our velocity system) 
and its error measured in \citet{sch06}, and their IDs, respectively.
The final column gives a merged, weighted mean velocity for all objects.
We have determined the membership of the objects in the master catalog as described in the following
section. 
The genuine NGC 4636 GCs
are listed first in the table, followed by eleven foreground stars,
two background galaxies and the NGC 4636 nucleus, and three probable
intracluster GCs of the Virgo cluster.

\subsection{Velocity Distribution and Membership}

We have determined the membership of the objects in the master catalog
using the distribution of radial velocity.
In Figure 5 %$ \ref{fig05}, 
we plot ($C-T_1$) colors versus the radial velocities, %the lower left panel),
  the radial velocity distribution, %the upper panels), 
  and the $(C-T_1)$ color distribution  %the lower right panel) 
  for the objects in the master catalog. 
The color distribution of the GCs in NGC 4636 is  bimodal,  which
  %through the KMM mixture modeling routine \citep{ash94}, and 
   is fit well by two Gaussians with peaks 
at %$(C-T_1)=1.30$ ($\sigma=0.20$) and 1.73 ($\sigma=0.22$)
$(C-T_1)=1.29$ ($\sigma=0.17$) and 1.76 ($\sigma=0.14$)
 \citep{lee03,dir05,par09}.
The minimum between the two components is at $(C-T_1)=1.55$, which was used
for diving the entire sample into the blue GCs and red GCs.

The radial velocity distribution of the 252 objects excluding the galaxies is approximately Gaussian
%The radial velocity distribution of the 252 objects is approximately Gaussian
centered around the radial velocity for the NGC 4636 nucleus, $v_p\approx 928$ km~s$^{-1}$, 
with a weak excess around the zero velocity.
Figure 5 shows that the radial velocity of the objects with larger than 
  that of NGC 4636 nucleus is distributed up to $\sim 1600$ \kms.
Considering a symmetric distribution for the radial velocity
of NGC 4636 GCs, we adopted 300 \kms as a lower limit for
the radial velocity of NGC 4636 GCs to avoid a contamination of the objects
with low velocity.
Finally we selected, as genuine NGC 4636 GCs,
those for which radial velocities are in the range of $300 \le v_p \le 1600~{\rm
km~s}^{-1} $ and ($C-T_1$) colors are in the range of $0.9 \le (C-T_1) < 2.1$.
The dashed-line box in Figure 5 represents the boundary for selecting GCs.
%adopted selection criteria for ($C-T_1$) colors and radial velocities.
The selection color range is equivalent to a metallicity range of
$-$2.24 $\lesssim$~[Fe/H]~$\lesssim$ 0.33 dex
%(or $-$2.45$\lesssim$~[Fe/H]~$\lesssim$0.41 dex),
%$-$2.37$\lesssim$~[Fe/H]~$\lesssim$0.05 dex, and
%$-$2.25$\lesssim$~[Fe/H]~$\lesssim$0.37 dex
using the color-metallicity relations in \citet{lee08b}.
% (double linear and 3rd order polynomial equations, respectively).
%,\citet{har02}, and \citet{coh03}, respectively.
The total number of the GCs in NGC 4636 in this catalog is 238, which includes
108 blue GCs  and 130 red GCs, respectively.

There are five objects that have radial velocities larger than the upper boundary
for the NGC 4636 GCs: IDs 644, 1192, 1293, g1, and g2. 
We searched for these objects in the NASA Extragalactic Database (NED), but found none.
Two (g1 and g2) of these with \vp~$\sim$ 26,000 \kms are background galaxies beyond the Virgo cluster.
% because of \vp $\sim$ 26,000 \kms. 2
The other three with \vp~$\sim2300$ \kms %do not belong to NGC 4636, but 
are probably the intracluster GCs of the Virgo cluster.
%, although they do not belong to NGC 4636.
%These are probably intracluster GCs in the Virgo cluster.
There are 11 objects with radial velocities smaller than the lower boundary for the NGC 4636 GCs. We consider these objects as foreground stars.

Among the 119 GC candidates with radial velocities derived in this study, 
105 objects are found to be  GCs in NGC 4636,
3 objects are found to be probable intracluster GCs belonging to the Virgo cluster,
%3 objects are found to be  GCs belonging to the Virgo cluster,
and 11 objects turned out to be foreground stars.
Therefore the success rate of photometric searching for GCs,
becomes about 90\%,
showing that the method to select the GC candidates using the $(C-T_1)$
color and the morphological classifier is very efficient 
(see \citealt{par09,lee08b} for more details).

\subsection{The Final Sample of Globular Clusters in NGC 4636}

The positions of the  GCs are marked in Figure 1.
These GCs are located in the range of projected galactocentric distance %agtnradius
23\arcsec~to 926\arcsec~ 
(corresponding to 1.63 to 65.75 kpc and 0.26 to 10.36 $R/R_{eff}$), most of which are 
within $10\arcmin$.
%(corresponding to 2.52 to 45.36 kpc and 0.33 to 6.00 $R/R_{eff}$).
The velocity dispersion %of the genuine globular clusters
%(the biweight scale of \citealt{bee90a}) is estimated
obtained using the biweight scale of \citet{bee90a} is estimated
to be $\sigma_p=231_{-17}^{+15}$ km~s$^{-1}$ for the 105 GCs measured in this study,
and $\sigma_p=225_{-9}^{+12}$ km~s$^{-1}$ for all 238 GCs in the master catalog.
The velocity dispersion of all 108 blue GCs is derived to be
$\sigma_p=251_{-12}^{+18}$ km~s$^{-1}$, which is slightly larger than that of
the 130 red GCs ($\sigma_p=205_{-13}^{+11}$ km~s$^{-1}$).
The mean value of the radial velocity derived 
using the biweight location of \citet{bee90a} %(the biweight location of \citealt{bee90a})
is $\overline{v_p}=932_{-22}^{+25}$ km~s$^{-1}$ for the GCs measured in this study,
and $\overline{v_p}=949_{-16}^{+13}$ km~s$^{-1}$ for all  GCs in the master catalog.
These values are very similar to the radial velocity for the NGC 4636 nucleus, 
$v_{\rm gal}=928\pm45$ km s$^{-1}$.

The GCs close to the color boundary ($(C-T1)=1.55$) used for dividing the blue
and red GCs will be overlapped by the red tail and the blue tail of the color distribution.
To check the effect of this problem in calculating the velocity disperion, 
we selected the blue GCs with $0.9<(C-T_1 )<1.4$
and the red GCs with $1.7<(C-T_1 )<2.1$, and calculated their velocity dispersion.
The velocity dispersions thus derived are
$\sigma_p=242_{-24}^{+18}$ km~s$^{-1}$ for the blue GCs, and
$\sigma_p=196_{-17}^{+15}$ km~s$^{-1}$ for the red GCs.
These values are consistent with those of the sub-samples simply divided in the above
($251_{-12}^{+18}$ km~s$^{-1}$, and $205_{-13}^{+11}$ km~s$^{-1}$, respectively).
Therefore the effect of the color tail contamination in selecting the sub-samples is minor.
   
Figure 6 %~\ref{fig06} 
displays the $T_1 - (C-T_1)$ color magnitude diagram and color
distribution for all the GCs and the brightest GCs in NGC 4636.
For comparison we also plot the point sources in the photometric sample given in \citet{par09}.
Most of the objects with $0.9\leq(C-T_1 ) <2.1$ in the photometric sample are probably GCs. 
The GCs in the master catalog have magnitudes $19.0 < T_1 < 22.7$,
belonging to the brightest population in NGC 4636, and the ranges of color and magnitude
of the GCs in this study and \citet{sch06} are similar. 
The color distribution of the GCs is clearly bimodal  in the spectroscopic sample 
like in the photometric sample.
It is noted that the color distribution of the brightest objects with $19.0 < T_1 < 20.2$ 
shows a dominant red peak and a much weaker blue peak. 
This was also seen in the case of NGC 1399 \citep{ost98,dir03}.
The velocity dispersion of these objects is $\sigma_p=204_{-18}^{+13}$ km~s$^{-1}$, 
which is very similar to that of the red GCs. 
The radial number density profile of these bright objects is also similar to that of the red GCs.
Therefore most of these brightest objects are considered to be genuine GCs in NGC 4636.
Details of the property of these GCs will be given in \citet{par09}.

%These bright GCs are also as concentrated as the red GCs in the radial distribution,
%as if it is formerly known from photometric results of NGC 1399 GCs by \citet{dir03}.
%In photometric study of NGC 1399 GC system, 
% the bright GCs also has a unimodal distribution and 
% are as concentrated as the red GCs, not the blue GCs (\citealt{dir03}).
%These may imply that an origin of the brightest GCs and the red GCs in galaxy is same.

In Figure 7 we plot radial velocities versus 
$(C-T_1)$ colors and $T_1$ magnitudes of the GCs of NGC 4636.
The mean value of the radial velocities for the GCs, represented by squares with errorbars,
shows little dependence  on the magnitude or color of the GCs.
However, the scatter in the velocity for the blue GCs is slightly larger than that for the red GCs.
Also we derived a mean velocity dispersion profile for all the GCs
as a function of $(C-T_1 )$ color, using a moving color bin (width of 0.2 mag and step of 0.05), as plotted in Figure 7(a).
The profile shows that the mean velocity dispersion decreases by $\approx50$ km s$^{-1}$ 
as the color increases (from blue to red).

In Figure 8 we display radial velocities versus  position angle $\Theta$ %[deg]
(measured to  east from the north)  and  projected galactocentric radius $R$ of the GCs. %[arcsec] 
The mean value and scatter of the radial velocities for the GCs change little
depending on  the position angle or galactocentric radius.
These two figures also show that the radial velocity distributions 
against color, magnitude, $\Theta$, and $R$ in this study and \citet{sch06} are similar.

Figure 9 shows the radial velocity distributions for all the GCs,
the blue GCs, and the red GCs of NGC 4636 in the master catalog.
It also displays the distributions of the bright GCs ($T_1<21$ mag) and
the faint GCs ($T_1 \ge 21$ mag). 
The radial velocity distribution for the blue GCs shows 
a larger velocity dispersion than that for the red GCs, as derived previously.
The radial velocity distributions for the brights and faint GCs look similar.
There appears to be a small peak at $v_p \sim 700$ km s$^{-1}$ in the case of  the red GCs.
We checked the significance of this peak using the KMM test \citep{ash94} and the
$I$ stastistic \citep{tea90}.
The KMM test showed that
the probability that the radial velocity distribution for the red GCs is bimodal is only 23 \%.
The $I$ value for the red GCs is calculated to be 0.972, which is smaller than the
critical value for rejecting the Gaussian hypothesis at the 90\% confidence level,
$I_{0.90}$ =1.040, showing that the distribution for the red GCs is Gaussian.
These show that the existence of the small peak at $v_p \sim 700$ km s$^{-1}$ is not significant.
It is noted that \citet{ric04} also concluded 
that the significance of a weak peak seen at $v_p \sim 1800$ km s$^{-1}$ in the velocity distribution of the GCs in NGC 1399 is doubtful.
A full dynamical analysis of the kinematics of the GC system of NGC 4636
using these data will be given in the companion paper \citep{lee09}.

\section{Summary}

We have presented a spectroscopic database of GCs in NGC 4636.
The spectroscopic targets were selected from the catalog of GC candidates derived 
from $C$ and $T_1$ images taken at the KPNO 4m\citep{par09}.
The spectra of those objects were obtained using MOS mode of FOCAS on the SUBARU Telescope.
We classified our targets into 105  GCs in NGC 4636, 11 stars, 
and 3 probable intracluster GCs in Virgo, using the $(C-T_1)$ color and the radial velocity.
Primary results in this study are summarized as follows.

\begin{enumerate}

 \item We measured the radial velocities for 122 objects in the field of NGC 4636:
105 GCs (51 blue GCs  %with $0.9\le(C-T_1)<1.55$ and
 and 54 red GCs), % with  $1.55\le(C-T_1)<2.1$), 
2 background galaxies, 3 probable intracluster GCs in  Virgo,  the nucleus of NGC 4636, and 11 foreground stars.

\item The mean value and dispersion of radial velocities for the 105 GCs measured in this study
are derived to be $\overline{v_p}=932_{-22}^{+25}$ km~s$^{-1}$ and $\sigma_p=231_{-17}^{+15}$ km~s$^{-1}$, respectively.

\item We combined our result with the data for 174 GCs of NGC 4636 given by \citet{sch06},
and created a master catalog of radial velocity for 255 objects in the field of NGC 4636.
The number of the GCs  in the master catalog is 238. 
The numbers of the blue GCs and red GCs are 108 and 130, respectively.

\item The mean value and dispersion of radial velocities for all  GCs
are derived to be 
$\overline{v_p}=949_{-16}^{+13}$ km~s$^{-1}$ and $\sigma_p=225_{-9}^{+12}$ km~s$^{-1}$, respectively.
The velocity dispersions of the blue GCs and red GCs
are estimated to be $\sigma_p=251_{-12}^{+18}$ km~s$^{-1}$ and $\sigma_p=205_{-13}^{+11}$ km~s$^{-1}$, respectively.

\item The mean radial velocities for the GCs change little depending on the magnitude, color,  position angle or galactocentric radius.

\end{enumerate}

\acknowledgments
The authors are grateful to the staff of the SUBARU Telescope for their kind help during the observation.
M.G.L. is supported in part by a grant
(R01-2007-000-20336-0) from the Basic Research Program of the
Korea Science and Engineering Foundation.
N.A. is financially supported in part by a Grant-in-Aid for
Scientific Research by the Japanese Ministry of Education, Culture,
Sports, Science and Technology (No. 19540245).

%{\it Facilities:} \facility{Nickel}, \facility{HST (STIS)}, \facility{CXO (ASIS)}.
%\appendix
%\section{Appendix material}

\clearpage
%\input{tabfigs3}
%\input{tabs}
%
%%%%%%%%%%%%%%%%%%%%%%%%%%%%%%%%%%%%%%%%%%%%%%%%%%%%%%%%%%%%%%%%%%%%%%%%%%%%%%%
%%%%%%%%%%%%%%%%%%%%%%%%         table 1       %%%%%%%%%%%%%%%%%%%%%%%%%%%%%%%%
%%%%%%%%%%%%%%%%%%%%%%%%%%%%%%%%%%%%%%%%%%%%%%%%%%%%%%%%%%%%%%%%%%%%%%%%%%%%%%%

%\begin{center}
 %\end{center}
\begin{deluxetable}{clccccccc}
\tabletypesize{\scriptsize}
\tablewidth{0pc}
\tablecaption{Observing Log for the Subaru FOCAS/MOS Run\label{tab-mask}}
\tablehead{
%\colhead{Obs. Date (UT)} &
\colhead{Mask Name} &
\colhead{R.A.(J2000)} &
\colhead{Decl.(J2000)} &
\colhead{N(objects)} &
\colhead{T(exp)} &
\colhead{seeing (\arcsec)} &
\colhead{Date(UT)}
}
\startdata
Mask-C & 12 42 50.0 & 2 40 47.4 & 36 & 2$\times$1200, 1800 s & 0.8 & Apr 20, 2002 \\
Mask-1 & 12 42 33.6 & 2 44 25.8 & 35 & 3$\times$1200 s & 0.8 & Apr 20, 2002 \\
Mask-3 & 12 43 08.2 & 2 43 23.4 & 34 & 3$\times$1200 s & 0.6 & Apr 21, 2002 \\
Mask-4 & 12 43 06.3 & 2 38 01.8 & 36 & 3$\times$1200 s & 0.6 & Apr 21, 2002 \\
Mask-6 & 12 43 03.7 & 2 48 49.8 & 27 & 4$\times$1200 s & 0.6 & Apr 21, 2002 \\
\enddata
\end{deluxetable}
\clearpage

%%%%%%%%%%%%%%%%%%%%%%%%%%%%%%%%%%%%%%%%%%%%%%%%%%%%%%%%%%%%%%%%%%%%%%%%%%%%%%%
%%%%%%%%%%%%%%%%%%%%%%%%         table 2       %%%%%%%%%%%%%%%%%%%%%%%%%%%%%%%%
%%%%%%%%%%%%%%%%%%%%%%%%%%%%%%%%%%%%%%%%%%%%%%%%%%%%%%%%%%%%%%%%%%%%%%%%%%%%%%%

\begin{deluxetable}{crrcc}
\tablewidth{0pc}
\tablecaption{A List of Galactic GCs Used as Templates \label{tab-galgc}}
\tablehead{
\colhead{Cluster} &
\colhead{[Fe/H]\tablenotemark{a}} &
\colhead{$v_{hel}$\tablenotemark{a}} &
\colhead{$E(B-V)$\tablenotemark{a}} &
\colhead{T(exp)}  \\
\colhead{} &
\colhead{(dex)} &
\colhead{(km s$^{-1}$)} &
\colhead{} &
\colhead{(sec)} 
}
\startdata
NGC 6341 (M92) & $-$2.28 & $-120.3\pm0.1$ & 0.02 & 300 \\
NGC 6205 (M13) & $-$1.54 & $-245.6\pm0.3$ & 0.02 & 300 \\
NGC 5904 (M5) & $-$1.27 &  $52.63\pm0.4$   & 0.03 & 240 \\
NGC 6171 (M107) & $-$1.04 &  $-33.6\pm0.3$   &0.33 & 300 \\
NGC 6624      & $-$0.44 &        $53.9\pm0.6$    &0.28 & 300 \\
\enddata
\tablenotetext{a~}{From \citet{har96}.}
\end{deluxetable}
\clearpage

%%%%%%%%%%%%%%%%%%%%%%%%%%%%%%%%%%%%%%%%%%%%%%%%%%%%%%%%%%%%%%%%%%%%%%%%%%%%%%%
%%%%%%%%%%%%%%%%%%%%%%%%         table 3       %%%%%%%%%%%%%%%%%%%%%%%%%%%%%%%%
%%%%%%%%%%%%%%%%%%%%%%%%%%%%%%%%%%%%%%%%%%%%%%%%%%%%%%%%%%%%%%%%%%%%%%%%%%%%%%%
\begin{deluxetable}{rrrrrrrrrrr}
\rotate
\tabletypesize{\scriptsize}
\tablewidth{0pc}
\tablecaption{Radial Velocities of the Objects in the Field of NGC 4636 \label{tab-m60gc}}
\tablehead{
\colhead{ID$^a$} &
\colhead{R.A.} &
\colhead{Decl.} &
\colhead{$R$} &
\colhead{$\Theta$} &
\colhead{$T_1$} &
\colhead{$(C-T_1)$} &
\colhead{$v_p$} &
\colhead{$v_p$(S06)} &
\colhead{ID(S06)} &
\colhead{$<v_p>$} \\
%\cline{9-10}
\colhead{} &
\colhead{(J2000)} &
\colhead{(J2000)} &
\colhead{(arcsec)} &
\colhead{(deg)} &
\colhead{(mag)} &
\colhead{(mag)} &
\colhead{(km s$^{-1}$)} &
\colhead{(km s$^{-1}$)} & &
\colhead{(km s$^{-1}$)}
}
\startdata

\multicolumn{11}{c}{}\\ \multicolumn{11}{c}{\underbar{Globular Clusters}}\\ \multicolumn{11}{c}{}\\
   627 & 12:42:51.95 &  2:41:58.5 &  53.3 &  36.6 & $21.21\pm0.03$ & $1.91\pm0.04$ & $ 1047\pm 69$ & $    ...    $ &   ...   & $ 1047\pm 69$ \\
   321 & 12:42:43.17 &  2:41:46.3 & 104.6 & 287.0 & $20.53\pm0.02$ & $1.64\pm0.03$ & $ 1078\pm 39$ & $ 1051\pm 19$ & 1.1-093 & $ 1056\pm 17$ \\
   566 & 12:42:55.43 &  2:41:43.5 &  88.3 &  71.7 & $21.09\pm0.02$ & $1.30\pm0.03$ & $  524\pm 52$ & $    ...    $ &   ...   & $  524\pm 52$ \\
..........\\
\multicolumn{11}{c}{}\\ \multicolumn{11}{c}{\underbar{Stars}}\\ \multicolumn{11}{c}{}\\
  1348 & 12:42:33.86 &  2:45:39.7 & 356.4 & 317.8 & $21.98\pm0.03$ & $1.66\pm0.05$ & $  200\pm 48$ & $    ...    $ &   ...   & $  200\pm 48$ \\
   204 & 12:42:35.24 &  2:41:45.4 & 220.7 & 277.7 & $20.09\pm0.02$ & $1.81\pm0.02$ & $   39\pm 28$ & $    ...    $ &   ...   & $   39\pm 28$ \\
    90 & 12:43:09.34 &  2:42:21.4 & 299.5 &  77.3 & $19.17\pm0.01$ & $1.27\pm0.01$ & $  -32\pm 22$ & $    ...    $ &   ...   & $  -32\pm 22$ \\
..........\\
\multicolumn{11}{c}{}\\ \multicolumn{11}{c}{\underbar{Galaxies}}\\ \multicolumn{11}{c}{}\\
    g1 & 12:42:38.39 &  2:46:10.2 & 340.8 & 329.8 & $ ... $ & $ ... $ & $26096\pm109$ & $    ...    $ &   ...   & $26096\pm109$ \\
    g2 & 12:43:02.44 &  2:50:11.4 & 568.0 &  19.4 & $ ... $ & $ ... $ & $25673\pm197$ & $    ...    $ &   ...   & $25673\pm197$ \\
NGC 4636 & 12:42:49.84 &  2:41:15.7 &  0.0 & 0.0 & $ ... $ & $ ... $ & $  928\pm 45$ & $    ...    $ &   ...   & $  928\pm 45$ \\
\multicolumn{11}{c}{}\\ \multicolumn{11}{c}{\underbar{Probable Intracluster Globular Clusters in the Virgo Cluster}}\\ \multicolumn{11}{c}{}\\
  1192 & 12:43:12.88 &  2:45:21.1 & 423.4 &  54.6 & $21.83\pm0.02$ & $1.28\pm0.04$ & $ 2355\pm 61$ & $    ...    $ &   ...   & $ 2355\pm 61$ \\ 
  1293 & 12:43:00.62 &  2:41:39.4 & 163.3 &  81.6 & $21.94\pm0.03$ & $1.90\pm0.05$ & $ 2088\pm 65$ & $    ...    $ &   ...   & $ 2088\pm 65$ \\ 
   644 & 12:43:15.71 &  2:45:07.0 & 451.3 &  59.2 & $21.23\pm0.01$ & $1.31\pm0.02$ & $ 2346\pm 32$ & $    ...    $ &   ...   & $ 2346\pm 32$ \\ 
\enddata
\tablenotetext{a~}{From Park et al. (2009).}
\tablecomments{The complete version of this table is in the electronic edition of
the Journal. The printed edition contains only a sample.}
\end{deluxetable}
\clearpage
%

%%%%%%%%%%%%%%%%%%%%%%%%%%%%%%%%%%%%%%%%%%%%%%%%%%%%%%%%%%%%%%%%%%%%%%%%%%%%%%%
%%%%%%%%%%%%%%%%%%%%%%%%         figure 1       %%%%%%%%%%%%%%%%%%%%%%%%%%%%%%%
%%%%%%%%%%%%%%%%%%%%%%%%%%%%%%%%%%%%%%%%%%%%%%%%%%%%%%%%%%%%%%%%%%%%%%%%%%%%%%%
%%\input{figcaps}
\begin{figure}
\plotone{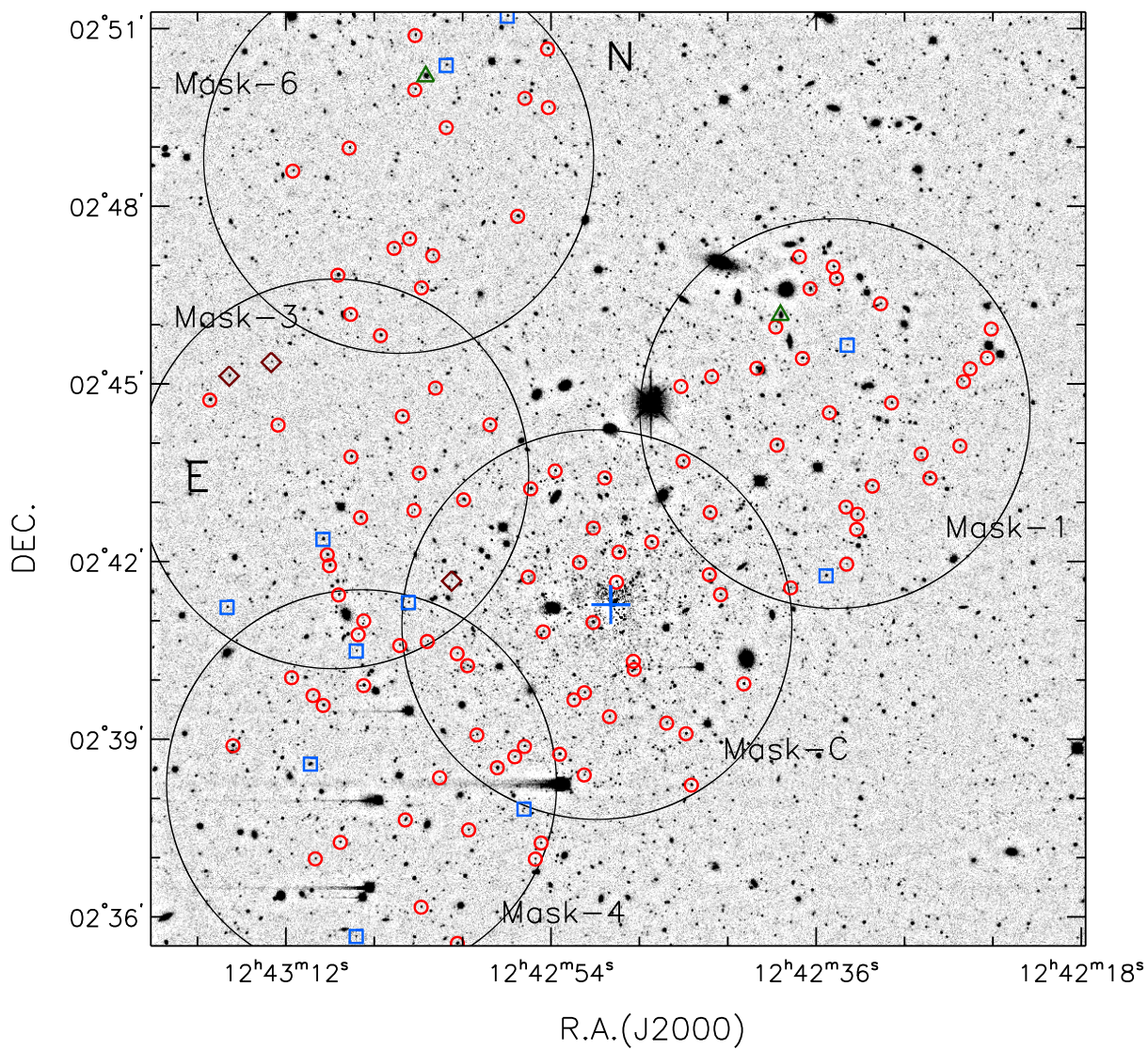}
\caption{
Positions of the observed masks (large circles with radius $3\arcmin$) overlaid on the grayscale map of
the Washington $T_1$ image ($15.8\arcmin \times 15.8\arcmin$)
with the spectroscopic sample of GC 
 candidates. North is up, and east to the left.
The galaxy light of NGC 4636 was subtracted from the original image using IRAF/ELLIPSE fitting.
The plus sign marks the center of NGC 4636.
Small circles, diamonds, triangles,   and squares represent the GCs in NGC 4636,
probable intracluster GCs in Virgo, background galaxies, and probable foreground stars, respectively.
\label{fig01}}%-mask}}
\end{figure}
\clearpage

%%%%%%%%%%%%%%%%%%%%%%%%%%%%%%%%%%%%%%%%%%%%%%%%%%%%%%%%%%%%%%%%%%%%%%%%%%%%%%%
%%%%%%%%%%%%%%%%%%%%%%%%         figure 2       %%%%%%%%%%%%%%%%%%%%%%%%%%%%%%%
%%%%%%%%%%%%%%%%%%%%%%%%%%%%%%%%%%%%%%%%%%%%%%%%%%%%%%%%%%%%%%%%%%%%%%%%%%%%%%%
\begin{figure}
\plotone{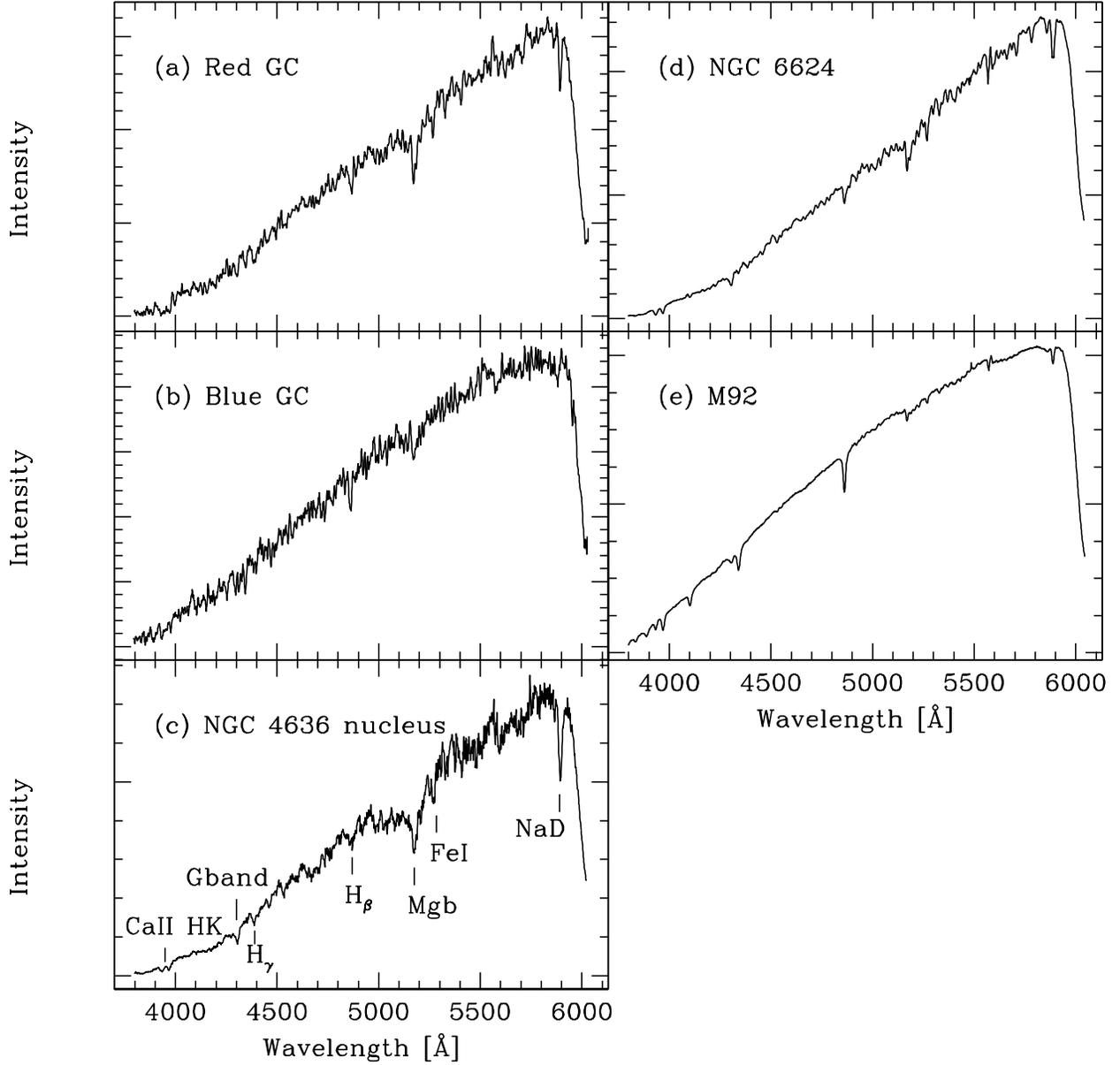}
\caption{
Example spectra of
(a) a red GC in NGC 4636 (ID: 154) with $T_1=19.82$ mag  and $(C-T_1)=1.73$;
(b) a blue GC in NGC 4636 (ID: 304) with $T_1=20.48$ mag and $(C-T_1)=1.31$;
(c) NGC 4636 nucleus;
(d) NGC 6624, a metal-rich Galactic GC with [Fe/H] = $-$0.44 dex; and
(e) M92, a metal-poor Galactic GC with [Fe/H] = $-$2.28 dex.
All spectra 
are plotted in the rest frame and the spectra of the blue and the red GCs are smoothed
using a boxcar filter with a size of 5 pixels.
\label{fig02}} %-exam}}
\end{figure}
\clearpage

%%%%%%%%%%%%%%%%%%%%%%%%%%%%%%%%%%%%%%%%%%%%%%%%%%%%%%%%%%%%%%%%%%%%%%%%%%%%%%%
%%%%%%%%%%%%%%%%%%%%%%%%         figure 3       %%%%%%%%%%%%%%%%%%%%%%%%%%%%%%%
%%%%%%%%%%%%%%%%%%%%%%%%%%%%%%%%%%%%%%%%%%%%%%%%%%%%%%%%%%%%%%%%%%%%%%%%%%%%%%%
\begin{figure}
%\epsscale{.80}
\plotone{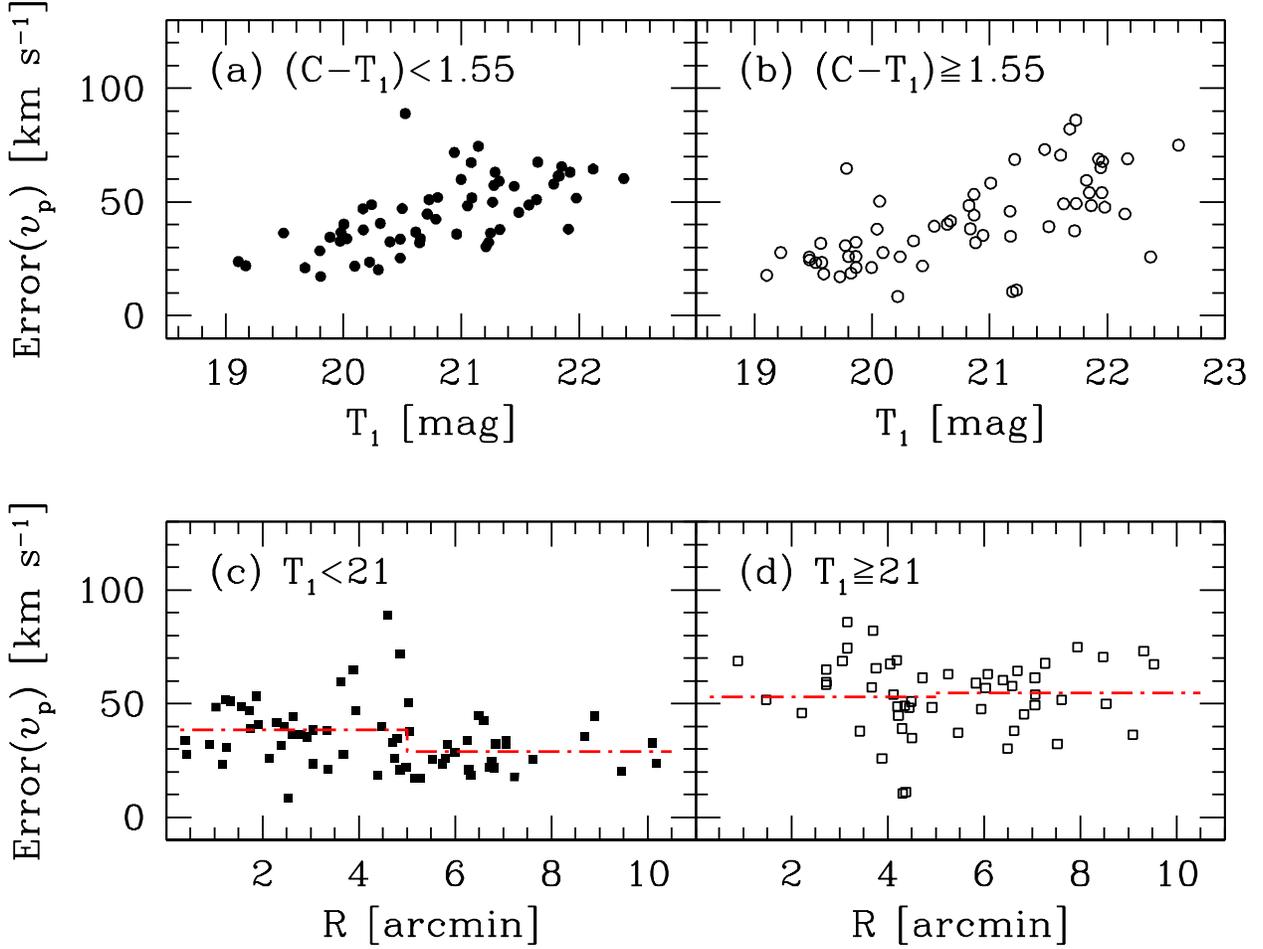}
\caption{
%{\it Upper panel}: 
Uncertainties of the measured radial velocities versus $T_1$ magnitudes (a and b) and versus galactocentric distances (c and d) for 119 spectroscopic targets in this study:
(a) the blue objects ($(C-T_1 ) <1.55$) (b) the red objects ($(C-T_1 ) \ge1.55$),
(c) the bright objects ($T_1<21$ mag), and (d) the faint objects ($T_1\ge21$ mag).
The dot-dashed lines are the mean uncertainties of the inner region ($R <5$ arcmin)
and outer region ($R \ge5$ arcmin).
\label{fig03}} %-verr}}
\end{figure}
\clearpage

%%%%%%%%%%%%%%%%%%%%%%%%%%%%%%%%%%%%%%%%%%%%%%%%%%%%%%%%%%%%%%%%%%%%%%%%%%%%%%%
%%%%%%%%%%%%%%%%%%%%%%%%         figure 4       %%%%%%%%%%%%%%%%%%%%%%%%%%%%%%%
%%%%%%%%%%%%%%%%%%%%%%%%%%%%%%%%%%%%%%%%%%%%%%%%%%%%%%%%%%%%%%%%%%%%%%%%%%%%%%%
\begin{figure}
\epsscale{.80}
\plotone{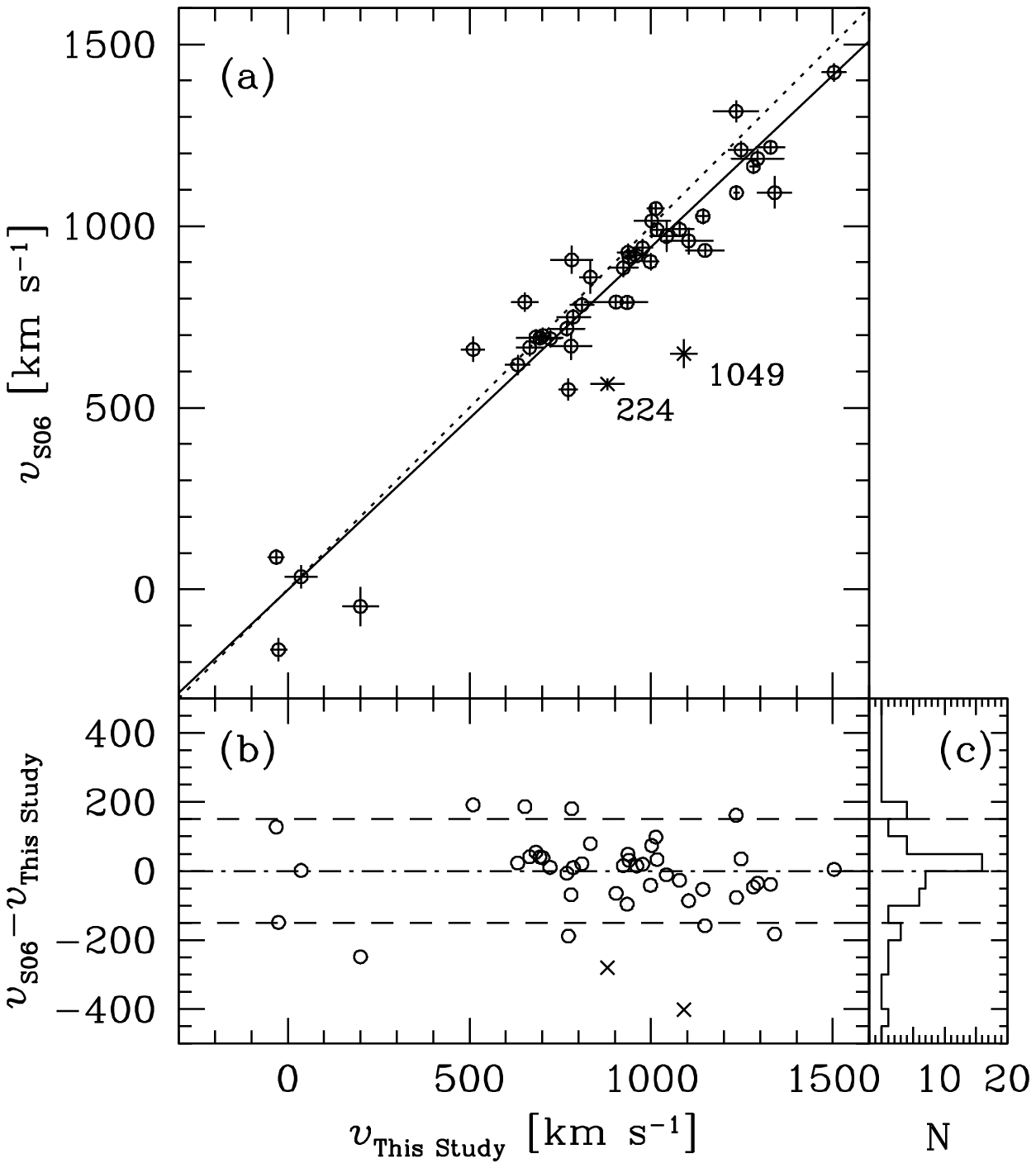} \caption{(a) Comparison of radial velocities of the objects
measured in this study and in \citet{sch06}. 
(b) The velocity difference ($v$(this study) minus $v$\citep{sch06}) versus $v$(this study).
(c) The distribution of the velocity difference.
The velocity uncertainty of each measurement is represented by horizontal and vertical bars.
The solid line represents the least-squares fit,
and the dotted line denotes the one-to-one relation.
Two objects marked by crosses were not used for the fit.
The dashed lines in lower panels indicate twice the $rms$ of the relation between the two studies.
\label{fig04}} %-vcomp}}
\end{figure}
\clearpage

%%%%%%%%%%%%%%%%%%%%%%%%%%%%%%%%%%%%%%%%%%%%%%%%%%%%%%%%%%%%%%%%%%%%%%%%%%%%%%%
%%%%%%%%%%%%%%%%%%%%%%%%         figure 5       %%%%%%%%%%%%%%%%%%%%%%%%%%%%%%%
%%%%%%%%%%%%%%%%%%%%%%%%%%%%%%%%%%%%%%%%%%%%%%%%%%%%%%%%%%%%%%%%%%%%%%%%%%%%%%%
\begin{figure}
\epsscale{.95}
\plotone{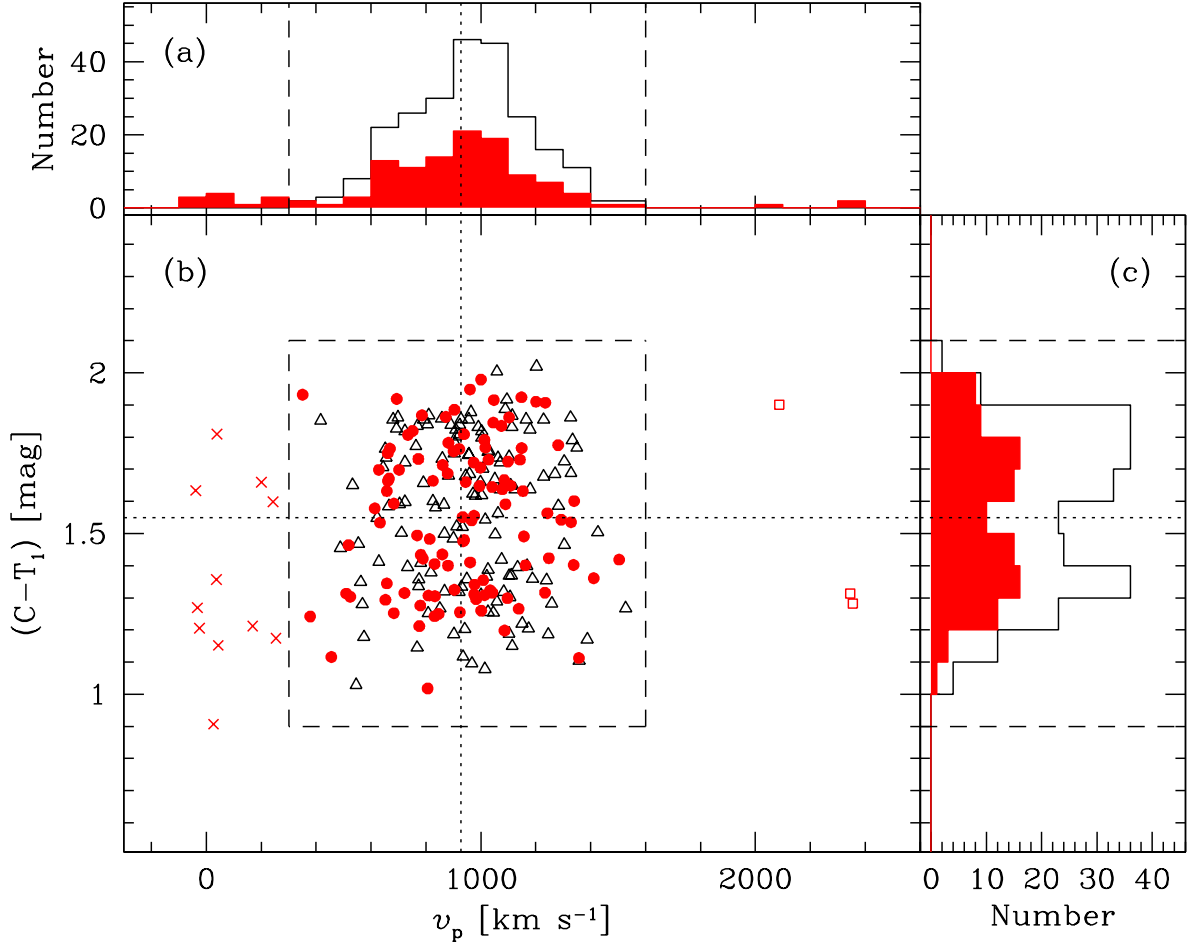} \caption{
(a) Radial velocity distribution, (b) $(C-T_1)$ colors versus radial
velocities, and (c) the $(C-T_1)$ color distribution for 105 GCs (filled circles) derived in this
study and 133 GCs (open triangles) measured only by \citet{sch06}.
The dashed box indicates the criteria
of radial velocities and $(C-T_1)$ colors for selecting GCs in NGC 4636.
Crosses near zero velocity are probably foreground stars,
and three objects represented by open squares are probable Virgo intracluster members.
The dotted vertical line indicates the velocity for the NGC 4636 nucleus, and the dotted
horizontal line denotes the color value that divides the GCs into
blue and red subsamples.
The shaded and open histograms in the radial velocity distribution and the color distribution 
represent the objects measured in this study, and all the objects in both this study and \citet{sch06}, respectively.
\label{fig05}}%-member}}
\end{figure}
\clearpage

%%%%%%%%%%%%%%%%%%%%%%%%%%%%%%%%%%%%%%%%%%%%%%%%%%%%%%%%%%%%%%%%%%%%%%%%%%%%%%%
%%%%%%%%%%%%%%%%%%%%%%%%         figure 6       %%%%%%%%%%%%%%%%%%%%%%%%%%%%%%%
%%%%%%%%%%%%%%%%%%%%%%%%%%%%%%%%%%%%%%%%%%%%%%%%%%%%%%%%%%%%%%%%%%%%%%%%%%%%%%%
\begin{figure}
\epsscale{0.8}
\plotone{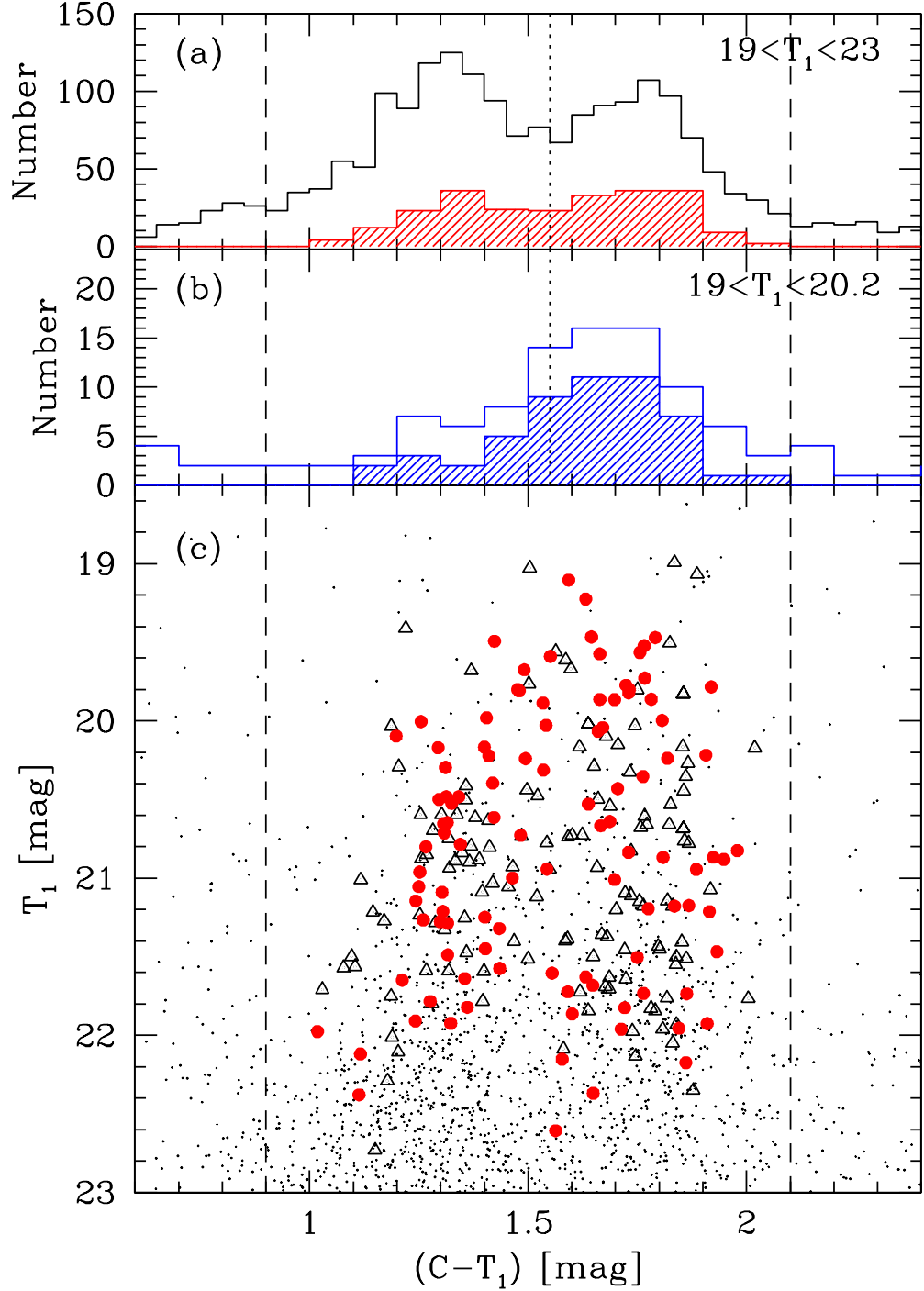}
\caption{%{\it Upper panel}: 
(a) $(C-T_1)$ color distribution for 238 GCs ($19<T<23$) (hatched histogram)
compared with that for the photometric sample of point sources (open histogram, \citealt{par09}). %GC candidates 
The dotted vertical line denotes the color value that divides the GCs into
the blue and the red subsamples,
and the dashed vertical lines denote $(C-T_1)$ color boundary for NGC 4636 GC candidates.
(b) Same as (a), but for the brightest GCs ($19<T<20.2$).
(c) $T_1 - (C-T_1)$ color magnitude diagram for 238 GCs in NGC 4636 from this study and \citet{sch06}. 
Filled circles indicate the
GCs measured in this study, and open triangles are the GCs measured only in
\citet{sch06}. 
Dots denote the photometric sample of point sources %NGC 4636 GC candidates 
at $R<720\arcsec$ \citep{par09}.
\label{fig06}}%-cmd}}
\end{figure}
\clearpage

%%%%%%%%%%%%%%%%%%%%%%%%%%%%%%%%%%%%%%%%%%%%%%%%%%%%%%%%%%%%%%%%%%%%%%%%%%%%%%%
%%%%%%%%%%%%%%%%%%%%%%%%         figure 7       %%%%%%%%%%%%%%%%%%%%%%%%%%%%%%%
%%%%%%%%%%%%%%%%%%%%%%%%%%%%%%%%%%%%%%%%%%%%%%%%%%%%%%%%%%%%%%%%%%%%%%%%%%%%%%%
\begin{figure}
\epsscale{1.0}
\plotone{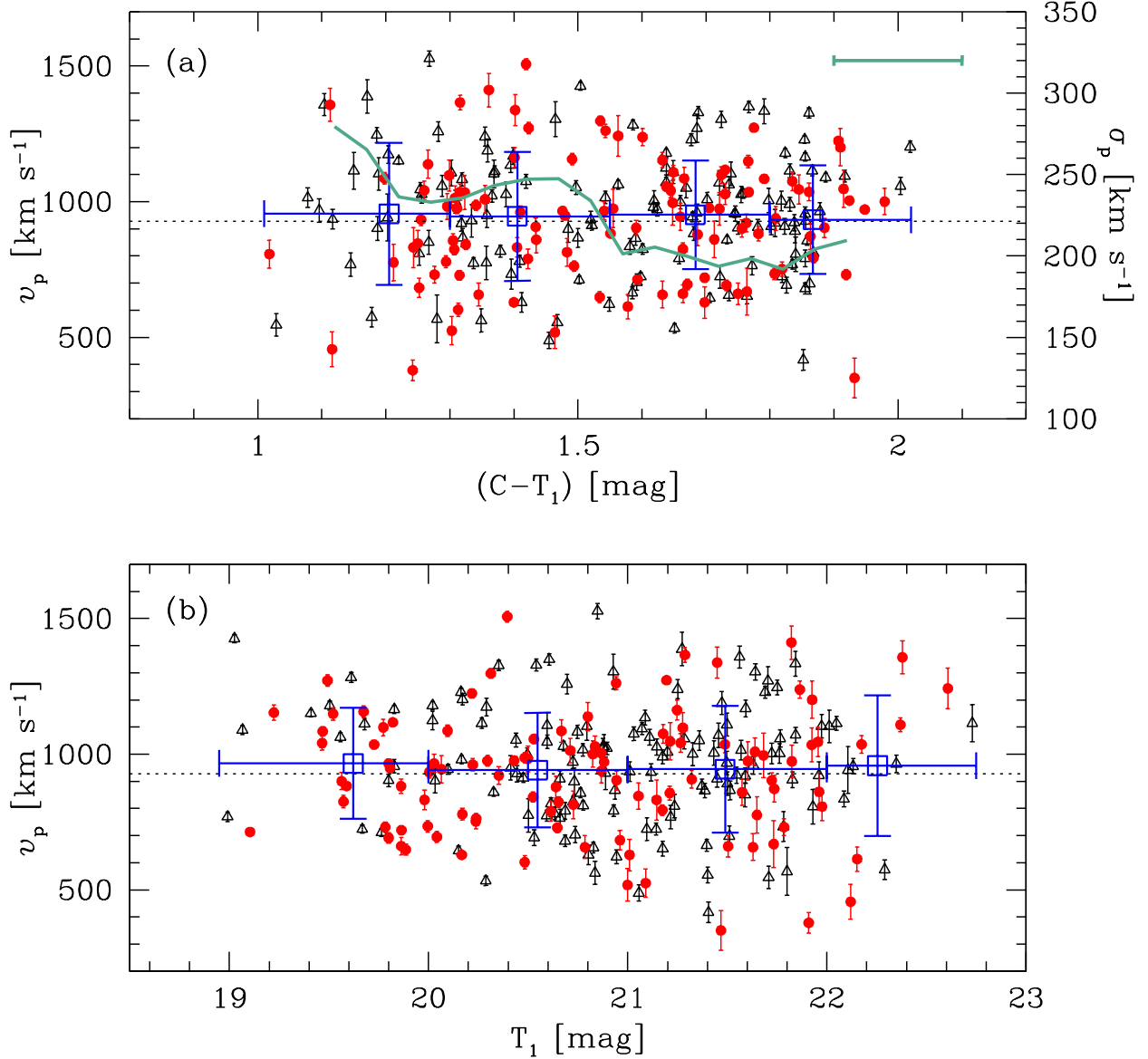} %fig7a.eps}
\caption{Radial velocities with measured errors versus $(C-T_1)$ colors (a), and versus 
$T_1$ magnitudes (b) for the GCs in NGC 4636.
Filled circles  and open triangles represent the GCs measured in this study
and only in \citet{sch06}, respectively.
%Open squares with errorbars represent the mean values.
Open squares indicate the mean radial velocities of the GCs in each bin, 
which is represented by a long horizontal error bar. 
Vertical error bars denote the velocity dispersions of GCs in the radial bins.
The dotted horizontal line indicates the systemic velocity of NGC 4636.
The solid line in (a) represents the mean velocity dispersion profile
for all the GCs derived using a moving color bin (width of 0.2 
and step of 0.05), 
shown by the horizontal bar in the upper right corner).
\label{fig07}}%-velmag}}
\end{figure}

\clearpage

%%%%%%%%%%%%%%%%%%%%%%%%%%%%%%%%%%%%%%%%%%%%%%%%%%%%%%%%%%%%%%%%%%%%%%%%%%%%%%%
%%%%%%%%%%%%%%%%%%%%%%%%         figure 8       %%%%%%%%%%%%%%%%%%%%%%%%%%%%%%%
%%%%%%%%%%%%%%%%%%%%%%%%%%%%%%%%%%%%%%%%%%%%%%%%%%%%%%%%%%%%%%%%%%%%%%%%%%%%%%%
\begin{figure}
\plotone{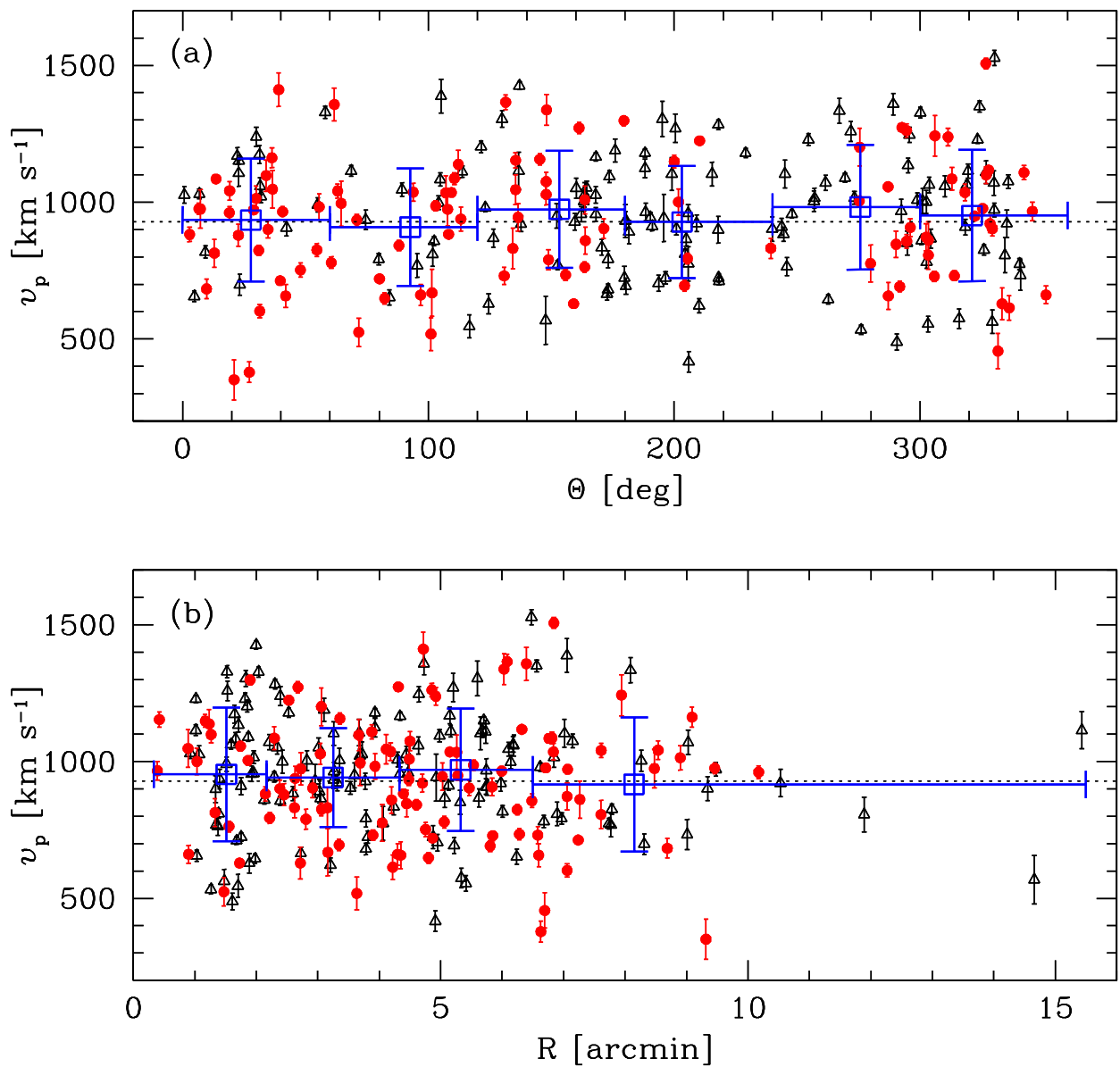} %fig8a.eps
\caption{Radial velocities with measured errors versus position angles (a), and versus
the projected galactocentric radii (b) for the GCs in NGC 4636.
The symbols are the same as in Fig. 7.
\label{fig08}}%-veldist}}
\end{figure}
\clearpage

%%%%%%%%%%%%%%%%%%%%%%%%%%%%%%%%%%%%%%%%%%%%%%%%%%%%%%%%%%%%%%%%%%%%%%%%%%%%%%%
%%%%%%%%%%%%%%%%%%%%%%%%         figure 9       %%%%%%%%%%%%%%%%%%%%%%%%%%%%%%%
%%%%%%%%%%%%%%%%%%%%%%%%%%%%%%%%%%%%%%%%%%%%%%%%%%%%%%%%%%%%%%%%%%%%%%%%%%%%%%%
\begin{figure}
\epsscale{0.7}
\plotone{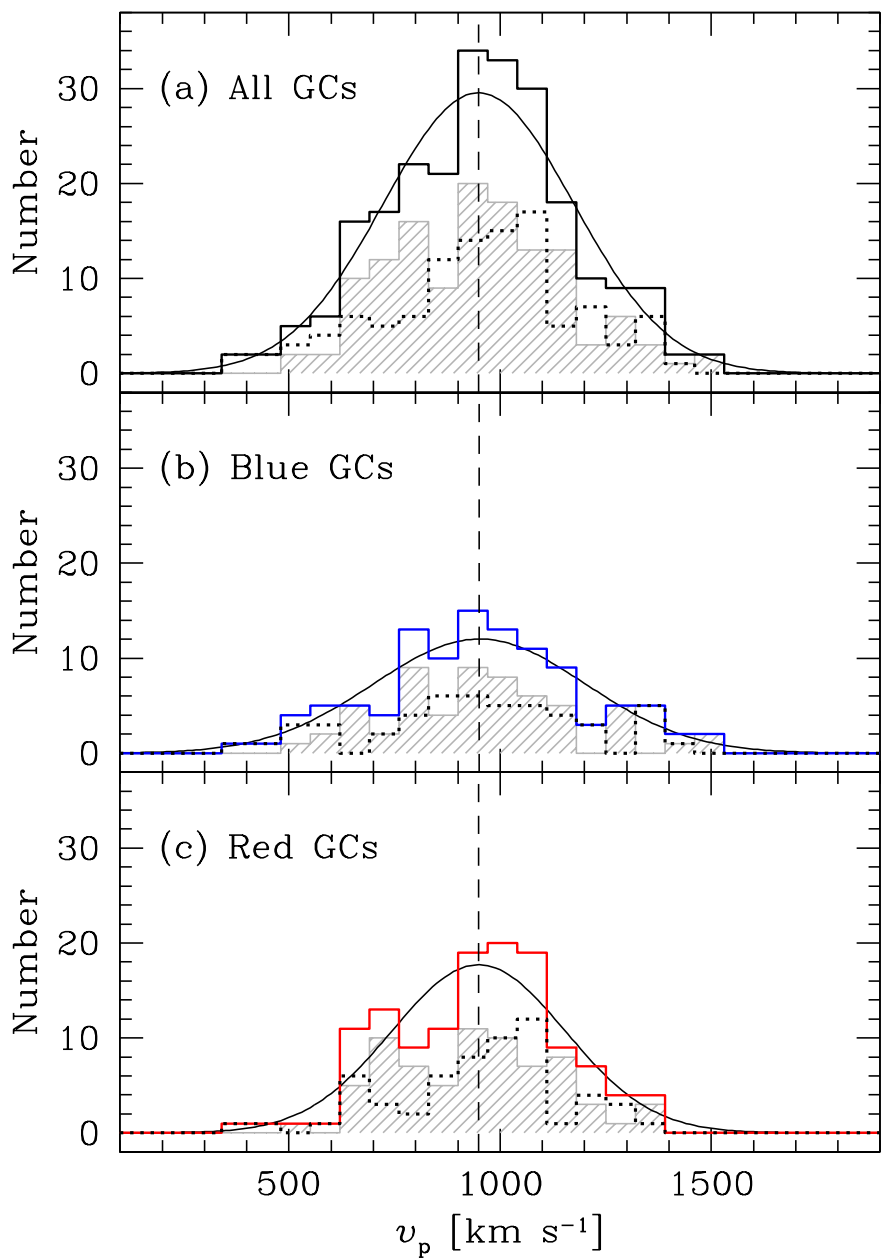}
\caption{
Radial velocity distribution of all the GCs (a), blue GCs (b), and red GCs (c) in NGC 4636.
The solid line, hashed, and dotted line histograms represent the distributions of the entire GCs,
the bright GCs ($T_1<21$ mag), and the faint GCs ($T_1 \ge 21$ mag), respectively.  
The solid curved lines represent Gaussian fits for the entire GCs
(with the mean velocity (the vertical dashed line) as a center, and the velocity dispersion as a width).
\label{fig09}}
\end{figure}
\clearpage


\begin{thebibliography}{}
\bibitem[Ashman et al.(1994)]{ash94} Ashman, K.~M., Bird, C.~M., \& Zepf, S.~E.\ 1994, \aj, 108, 2348 
\bibitem[Beasley \etal (2000)]{bea00} Beasley, M. A., Sharples, R. M., Bridges, T. J., Hanes, D. A., Zepf, S. E., Ashman, K. M.,
 \& Geisler, D. 2000, \mnras, 318, 1249 % N4472
\bibitem[Beers et al.(1990)]{bee90a} Beers, T. C., Flynn, K., \& Gebhardt, K. 1990, \aj, 100, 32
\bibitem[Bridges et al.(2006)]{bri06}Bridges, T., et al. 2006, \mnras, 373, 157
\bibitem[Brodie \& Hanes(1986)]{bro86} Brodie, J. P., \& Hanes, D. A. 1986, \apj, 300, 258 % MW GCs
\bibitem[Brodie \& Strader(2006)]{bro06} Brodie, J. P., \& Strader, J. 2006, \araa, 44, 193
\bibitem[Caon \etal (2000)]{cao00}Caon, N., Macchetto, D., \& Pastoriza, M. 2000, \apjs, 127, 39
\bibitem[Chakrabarty \& Raychaudury (2008)]{cha08} Chakrabarty, D. \& Raychaudury, S. 2008, \aj, 135, 2350 %DM n NGC 4636
\bibitem[Cohen \& Ryzhov(1997)]{coh97} Cohen, J. G., \& Ryzhov, A. 1997, \apj, 486, 230
\bibitem[Cohen \etal (1998)]{coh98} Cohen, J. G., Blakeslee, J. P., \& Ryzhov, A. 1998, \apj, 496, 808 %M87
\bibitem[C\^ot\'e \etal (2001)]{cot01} C\^ot\'e, P., \etal %McLaughlin, D. E., Hanes, D. A.,
%Bridges, T. J., Geisler, D., Merritt, D., Hesser, J. E., Harris, G. L. H., \& Lee, M. G. 
 2001, \apj, 559, 828
\bibitem[C\^ot\'e \etal (2003)]{cot03} C\^ot\'e, P., McLaughlin D. E., Cohen J. G., \& Blakeslee J. P. 2003, \apj, 591, 850 %M49
\bibitem[Dirsch et al. (2003)]{dir03} Dirsch, B., Richtler, T., Geisler, D., Forte, J. C., Bassino, L. P., \& Gieren, W. P.
 2003, \aj, 125, 1908 
\bibitem[Dirsch et al. (2005)]{dir05} Dirsch, B., Schuberth, Y., \& Richtler, T. 2005, \aap, 433, 43
\bibitem[Forman \etal (1985)]{for85}Forman, W., Jones, C., \& Tucker, W. 1985, \apj, 293, 102
%\bibitem[Hanes (1977a)]{han77a}Hanes, D. A. 1977a, \mnras, 180, 309 %N4636
%\bibitem[Hanes (1977b)]{han77b}Hanes, D. A. 1977b, MemRAS, 84, 45 %N4636
\bibitem[Harris (1996)]{har96} Harris, W. E. 1996, \aj, 112, 1487 (February 2003 version)
\bibitem[Hwang \etal (2008)]{hwa08}Hwang, H. S., \etal %Lee, M. G., Park, H. S., Kim, S. C., Sohn, Y.-J., Park, J. H.,   Lee, S.-G., Rey, S.-C., Lee, Y.-W., Kim. H. 
 2008, \apj, 674, 869
\bibitem[Jones et al. (2002)]{jon02}Jones, C., Forman, W., Vikhlinin, A., Markevitch, M., David, L, Warmflash, A., \& Murray, S. 2002, \apj, 567, L115 %CXO NGC 4636
\bibitem[Kashikawa et al. (2002)]{kas02} Kashikawa, N., et al. %Aoki, K., Asai, R., \& 16 coauthors. 
 2002, \pasp, 54, 819
\bibitem[Kim et al. (2006)]{kim06} Kim, E., Kim, D.-W., Fabbiano, G., Lee, M. G., Park, H. S., \& Dirsch, B. 2006, \apj, 647, 276
\bibitem[Kissler et al. (1994)]{kis94}Kissler, M., Richtler, T., Held, E. V., Grebel, E. K., Wagner, S. J., \& Capaccioli, M. 1994, \aap, 287, 463 % NGC4636  GCs
\bibitem[Kissler-Patig et al.(1998)]{kis98} Kissler-Patig, M., Brodie, J. P.,
Schroder, L. L., Forbes, D. A., Grillmair, C. J., \& Huchra, J. P. 1998, \aj, 115, 105
\bibitem[Kissler-Patig \& Gebhardt(1998)]{kis98b} Kissler-Patig, M., \&
Gebhardt, K.  1998, \aj, 116, 2237
\bibitem[Kissler-Patig et al.(1999)]{kis99} Kissler-Patig M., Grillmair C. J.,
Meylan G., Brodie, J. P., Minniti, D., \& Goudfrooij, P. 1999, \aj, 117, 1206
\bibitem[Lee (2003)]{lee03} Lee, M. G. 2003, Jour. Korean Astron. Soc.,  36, 189
\bibitem[Lee \etal(2008a)]{lee08a}Lee, M. G., \etal %Hwang, H. S., Park, H. S., Park, J. H., Kim, S. C. Sohn, Y.-J., Lee, S.-G., Rey, S.-C., Lee, Y.-W., Kim. H. 
 2008a, \apj, 674, 857
\bibitem[Lee \etal(2008b)]{lee08b}Lee, M. G., Park, H. S., Kim, E., Hwang, H. S., Kim, S. C., \& Geisler, D. 2008b, \apj, 682, 135 %M60 phot
\bibitem[Lee et al.(2009)]{lee09} Lee, M. G., Park, H. S.,  Hwang, H. S., Arimoto, N.,
Tamura, N., \& Onodera, M. 2009,  \apj, submitted (Paper II)
\bibitem[Loewenstein \& Mushotzky (2003)]{loe03}Loewenstein, M., \& Mushotzky, R. F. 2003, Nucl. Phys. B. Proc. Suppl., 124, 91
\bibitem[Matsushita et al. (1998)]{mat98}Matsushita, K., Makishima, K., Ikebe, Y., Rokunanda, E., Uamasaki, N. Y., \& Ohashi, T. 1998, \apj, 499, L13 % X-ray NGC 4636
\bibitem[Minniti et al.(1998)]{min98} Minniti, D., Kissler-Patig, M., Goudfrooij, P.,
\& Meylan, G.  1998, \aj, 115, 121
\bibitem[Nolthenius (1993)]{nol93}Nolthenius, R. 1993, \apjs, 85, 1 %Group catalog
\bibitem[Ostrov \etal (1998)]{ost98}Ostrov, P. G., Forte. J. C., \& Geisler, D. 1998, \aj, 116, 2854 
\bibitem[O'Sullivan et al. (2005)]{osu05}O'Sullivan, E., Vrtilek, J. M., \& Kempner, J. C. 2005, \apj, 624, L77 %CXO NGC4636
\bibitem[Park et al.(2009)]{par09} Park, H. S., Lee, M. G., et al. 2009, % Kim, E., Kim, S. C.      \& Geisler, D. 2009b, 
in preparation % NGC4636 phot
\bibitem[Peng et al.(2004a)]{pen04a} Peng, E. W., Ford, H. C., \&  Freeman, K. C. 2004a, \apjs, 150, 367 %NGC 5128 I
\bibitem[Peng et al.(2004b)]{pen04b} Peng, E. W., Ford, H. C., \&  Freeman, K. C. 2004b, \apj, 602, 705  %NGC 5128 II
\bibitem[Pierce et al.(2006)]{pie06} Pierce, M., et al. 2006, \mnras, 368, 325
\bibitem[Posson-Brown et al. (2009)]{pos09} Posson-Brown, J., Raychaudhury, S., Forman, W., Hank D. R., \& Jones, C. 2009, \apj, 695, 1094 %accepted by \apj, astro-ph/0605308 % LMXB vs GCs NGC 4636
\bibitem[Richtler et al.(2004)]{ric04} Richtler, T., et al. 2004, \aj, 127, 2094
\bibitem[Richtler et al. (2008)]{ric08}Richtler, T., Schuberth, Y., Hilker, M., Dirsch, B., Bassino, L.,
\& Romanowsky, A. J. 2008, \aap, 478, L23 %NGC 1399 DM or Mond?
\bibitem[Romanowsky et al. (2009)]{rom09} Romanowsky, A. J., Strader, J., Spitler, L. R ., Johnson, R., Brodie, J. P., Forbes, D. A., \& Ponman, T. 2009, \aj, 137, 4956 %NGC 1407
\bibitem[Smith et al. (2000)]{smi00}Smith, R. J., Lucey, J. R., Hudson, M. J., Schlegel, D. J., \& Davies, R. L. 2000, \mnras, 313, 469 %v(n4636)
\bibitem[Saito et al. (2003)]{sai03} Saito, Y., et al. 2003, Proc. SPIE, 4841, 1180
\bibitem[Schlegel et al.(1998)]{sch98} Schlegel, D. J., Finkbeiner,
D. P., \& Davis, M. 1998, \apj, 500, 525
\bibitem[Schuberth et al. (2006)]{sch06}Schubert, Y., Richtler, T., Dirsch, B., Hilker, M., Larsen, S. S., Kissler-Patig, M., \& Mebold, U. 2006, \aap, 459, 391 %in press (astroph/0604309)
\bibitem[Teague \etal (1990)]{tea90}Teague, P. F., Carter, D., \& Gray, P. M. 1990, \apjs, 72, 715 
%\bibitem[Temi et al. (2003)]{tem03}Temi, P., Mathews, W. G., Brighenti, F., \& Bregman, J. 2003, \apj, 585, L121 % ISO 60, 90, 180 um detection of NGC 4636
%\bibitem[Temi et al. (2007)]{tem07}Temi, P., Brighenti, F., \& Mathews, W. G. 2007, \apj, 666, 222 % 
\bibitem[Tonry \& Davis(1979)]{ton79} Tonry, J., \& Davis, M. 1979, \aj, 82, 954
\bibitem[Tonry et al. (2001)]{ton01} Tonry, J., et al. 2001, \apj, 546, 681 %SBF distance 
\bibitem[Woodley et al. (2007)]{woo07}Woodley, K. A., Harris, W. E., Beasley, M. A.,
Peng, E. W., Bridges, T. J., Forbes, D. A., \& Harris, G. L. H. 2007, \apj, 134, 494 %N5128
\bibitem[Zepf et al. (2000)]{zep00}Zepf, S. E., Beasley, M. A., Bridges, T. J., Hanes, D. A., Sharples, R. M., Ashman, K. M., \& Geisler, D. 2000, \aj, 120, 2928

\end{thebibliography}
\end{document}